\documentstyle[12pt,aaspp4]{article}

\def\day{{}$^{\rm d}$\llap{.}}

\accepted{August 22, 1997}

\begin{document}

\title{Evolution of Horizontal Branch Stars in Globular Clusters: \\
The Interesting Case of V79 in M3}

\author{\sc Christine M. Clement}

\affil{Department of Astronomy, University of Toronto \\
Toronto, Ontario, M5S 3H8, Canada\\
electronic mail: cclement@astro.utoronto.ca,}

\authoraddr{David Dunlap Observatory, University of Toronto \\
Toronto, Ontario, M5S 3H8, CANADA }

\author{\sc R. W. Hilditch}

\affil{School of Physics and Astronomy, University of St. Andrews \\
North Haugh, St. Andrews, Fife, KY16 9SS, Scotland, U.~K. \\
electronic mail:rwh@st-andrews.ac.uk}

\authoraddr{}

\author{\sc J. Kaluzny}

\affil{Warsaw University Observatory, Al. Ujazdowskie 4 \\
00-478 Warsaw, Poland\\
electronic mail: jka@sirius.astrouw.edu.pl}

\authoraddr{}

\author{\sc Slavek M. Rucinski\altaffilmark{1}}
 
\affil{81 Longbow Square, Scarborough, Ontario M1W~2W6, Canada \\
 electronic-mail: {rucinski@astro.utoronto.ca\/}}

\altaffiltext{1}{Affiliated with the Department of Astronomy,
University of Toronto and Department of Physics and Astronomy,
York University}

\begin{abstract}

New observations of variable stars in the globular cluster M3 
reveal that the RR Lyrae variable V79 is a double-mode (RRd) variable
with the first overtone mode dominating.
In all previous studies, V79 was found to be a fundamental mode (RRab)
pulsator with an irregular light curve. This  is the first observed mode 
switch for an RR Lyrae variable and it 
is direct observational evidence for blueward evolution of horizontal 
branch stars in the Oosterhoff type I cluster M3. 
It also demonstrates that there is a connection between the Blazhko 
effect and pulsational mode mixing in RR Lyrae variables.
These new observations also show that the strength of the overtone 
oscillations in
the RRd star V68 in M3 may have increased in the last 70 years, thus
indicating blueward evolution for V68 as well.

A survey of previously published investigations of RRd stars in Oosterhoff 
type II systems indicates that there is  marginal evidence for an increase 
in the strength of fundamental
mode oscillations in two stars: V30 in M15 and AQ Leo. If these increases
are confirmed by future observations, it will indicate redward 
evolution for RRd stars in type II systems.

\end{abstract} 
%
%************************ SECTION 1
%
\section{Introduction}
A standard method for investigating the evolution of horizontal branch
stars is to analyze the period changes of RR Lyrae variables in globular 
clusters.
Period increases should indicate redward evolution, while
decreases indicate
blueward evolution, and the rates of change give information about
the time scales.   
The models of Sweigert \& Renzini (1979) and
Lee \it et al. \rm (1990) predict that the RR Lyrae variables 
in the Oosterhoff type I clusters 
cross the instability strip during their
ZAHB phase and are therefore expected to have period decreases,
which are followed later by increases, when the stars evolve away
from the ZAHB. For the RR Lyrae variables in the more metal-poor
Oosterhoff type II
clusters, the models predict only increasing periods, because these stars
cross the instability strip after their ZAHB phase. 
Since there are a few globular clusters that have been observed over
an interval of 100 years, it should be possible to use these observations
to test the models.
Unfortunately, the observational data  do not give definitive results 
(cf. Smith 1995, Rathbun \& Smith 1997). 
The evolutionary period changes seem to be
masked by a period change $noise$ of irregular character. 
Sweigert \& Renzini (1979) have demonstrated that this $noise$ could be caused
by mixing events that alter the hydrostatic structure of the core and thus
affect the star's pulsation period. 
Because of this, the observed rates of change are sometimes an order
of magnitude too large and of the wrong sign, when compared with the
rates predicted by evolutionary theory. 
However, if one assumes that, for a particular cluster,
the $mean$ rate of period change 
is a measure of the evolutionary change, then it may be possible to
compare the observations with theory. Lee (1991) took this approach and 
found that the observed period changes of the RR Lyrae stars in five well
observed clusters could be attributed to evolutionary effects, provided
the $noise$ is random and of the order of $0.07$ days per million years.

Another way to study the evolution of HB stars 
is to examine the distribution of periods and modes of pulsation of RR
Lyrae variables in globular clusters of the two Oosterhoff types. Some years 
ago,
van Albada \& Baker (1973) postulated that the difference in the period
distributions for the Oosterhoff type I and II clusters 
is due in part to hysteresis in pulsation. 
According to their theory, the RRab variables in the type I clusters 
enter the instability strip as fundamental mode pulsators (RRab stars)
and evolve from red to blue, while  the stars in type II clusters enter
as first overtone pulsators (RRc stars) and evolve from blue to red. 
Because of hysteresis, the transition
period at which a star switches between the fundamental and first overtone
modes will be different for the two cluster types, and consequently,
the mean periods of both the RRc and RRab stars will be affected as well.
The RR Lyrae variables in type II clusters should switch modes at
longer periods and as a result, their mean periods will be longer
than those in the
type I clusters. Also, the Oosterhoff type II clusters 
should have a higher proportion
of RRc stars.
These are the observed characteristics of the two Oosterhoff groups.
If the van Albada \& Baker
scenario is correct, then the transition between the $ab-$ and 
$c-$type RR Lyrae variables should occur near the blue edge for
fundamental mode pulsation in Oosterhoff type I clusters and near the red
edge for the first overtone in Oosterhoff type II clusters. 
Using up-to-date convective
pulsation models to locate the instability strip on the HR diagram,
Bono \it et al.  \rm (1994) and Cox (1995) have
presented evidence to indicate that this is, in
fact, the case.

In the present investigation, we suggest another method for
studying HB evolution: monitoring long term
changes in the pulsation characteristics of double-mode RR Lyrae (RRd) stars.
If an RRd star is evolving blueward, then over a period of time, the amplitude
of the first overtone should gain in strength relative to fundamental, but 
if it evolves redward, the strength of fundamental mode oscillations should 
increase. Our study is motivated by the recent discovery of 
Kaluzny \it et al. \rm (1997, hereafter KHCR) that 
V79 in M3, previously classified as RRab, is now an RRd star with the 
first overtone mode dominating. 
The KHCR finding was based on observations made in 1996.

%
%************************ SECTION 2
%
\section{Analysis of V68 and V79 in M3}

The 1996 observations of M3 were obtained on nine nights in the interval
March 19 to April 2 by one of us (RWH)
with the 1 meter Jacobus Kapteyn telescope at the Observatorio del Roque de
Los Muchachos, La Palma. The observations, which include $V$ photometry for
42 RR Lyrae variables on 176 frames, and the reduction procedures,
have been discussed by KHCR. One
of the two previously known RRd variables,  V68, was among the 42 stars and so
we have included it in our investigation.
(V87, the other RRd star was not in our field of view.)
A finding chart for the variables in M3 was published by Bailey (1913).
We determined the
primary periods for V68 and V79 with  a computer program that
utilized Stellingwerf's (1978) phase dispersion minimization (PDM) technique
with a (5,2) bin structure. To search for the secondary period, we
derived a mean light curve by fitting
a cubic spline interpolating function to the bin means, then measured the 
residuals from this curve and applied the PDM technique 
to the residuals. Next, we corrected the magnitudes by  
subtracting the mean curve for the secondary period from the raw
magnitudes and then, again applied the PDM technique 
to obtain a final value for the primary period. 
Light curves for V68 and V79 are shown in Figures 1 and 2. The top
panel of each figure shows  the `raw' magnitudes plotted with the
primary period, the first overtone. 
The curves in the middle and bottom panels show respectively,
the corrected magnitudes plotted with the first overtone period
and the residuals plotted with the secondary (fundamental)
period, both with the interaction frequencies ($1/P_1 \pm 1/P_0$) also
removed.

To assess the long term behavior of these stars, we compared our
results with those of an earlier study
of the RRd stars in M3 by Nemec \& Clement (1989, hereafter
NC). The NC study was based mainly on data from a combination of 
three sets of observations made in  
the interval 1920 to 1926 and published by Larink (1922), Muller (1933)
and Greenstein (1935).  
NC found that V68 was an RRd star with the fundamental 
mode dominant, but they considered V79 to be an RRab star. 
The light curve of V79 (see Figure 2 of NC) showed night-to-night 
scatter, but a
PDM period search of the residuals did not reveal any periodicity in
the range expected for first overtone oscillations (see Figure 3 of NC). 
They estimated that if there
were any first overtone oscillations, they would have an amplitude less
than $0.25$ mag.

We summarize the pulsation characteristics of V68 and V79 
in Table 1.
In columns 2 and 3, we list their co-ordinates in arcseconds relative to the
cluster center according to Sawyer Hogg's (1973) catalogue, in columns
4 to 8, we list the first overtone
and fundamental periods ($P_1$ and $P_0$), their corresponding
amplitudes ($A_1$ and $A_0$) and amplitude ratios derived from the 1996 
observations and in  
the final column, we list
the amplitude ratios for 1920--1926 based on NC's study. 
The data of Table 1 illustrate that for
both stars, the first overtone oscillations have grown in strength
since the 1920s, but for V79, the change is more striking.

In order to find out what happened to V79 in the intervening years, we
used the PDM technique to
analyze other published observations that were suitable for period searches. 
These included observations
obtained in the intervals 1938 to 1962 (Szeidl 1965) and 1946 to
1948 (Belserene 1952). 
In the PDM technique, a $\Theta$ statistic which is 
a measure of the scatter on the light curve is evaluated for a series of 
periods and
the period for which $\Theta$ is a minimum is considered to be the best period.
In Figure 3 we
show plots of $\Theta$ versus period for the raw
data over the range 0\day 34 to 0\day 50, for the four epochs.
The plot for the 1996 data in the bottom panel of the figure indicates that
the best period is 0\day 358. However, the plots
for the first three epochs all show a 
clear minimum at a period of approximately 0\day 4833 and period searches of 
the residuals measured
from their corresponding  primary light curves did not reveal any oscillations 
in the overtone mode even though
there were variations in the amplitude.
Our analysis also indicates that the fundamental period of 
V79 has decreased by a substantial amount in the last 35 years.
The $0.4833$ day period derived from the previous studies is 
significantly longer than 
the fundamental  period $P_0=$0\day 480 that we determined for the 1996 
observations and listed in Table 1. If we plot the 1996
data with the longer period, it introduces a phase shift 
that increases the scatter on the light curve in the bottom panel of Figure 2.
We therefore conclude that two things have happened
to V79 between 1962 and 1996. First overtone oscillations have either
commenced or increased significantly and there  
has been an abrupt decrease in the fundamental period. 
In his study of period changes of the RR Lyrae variables in M3,
Szeidl (1965) found that the $\rm O - \rm C$ diagram for V79
had a discontinuity between 1926 and 1938 and
consequently, he was unable to detect any
systematic period change for the star. 
Rathbun \& Smith (1997) cited V79 as an example of a star that has ``period 
changes so
erratic as to be impossible to even approximately describe with a single
number.''
If the large abrupt decrease we have
detected in $P_0$ the fundamental period of V79 was
caused by evolution on the horizontal branch, then an $\rm O- \rm C$ diagram
is not a useful tool for studying evolutionary period changes of RR Lyrae
variables, at least not for stars like V79 in M3. 

Since, prior to 1962, V79 was an RRab star with an
irregular light curve: a `Blazhko' variable, 
our investigation 
provides some insight into the actual cause of the Blazhko effect in RR Lyrae
variables.
According to Smith (1995), most speculation has
centered on two possibilities: (1) that the effect is a ``consequence
of some type of mixing of pulsational modes'' and (2) that the
effect is ``related to magnetic cycles in the stars, perhaps coupled
with rotation''. Since we now know
that at least one `Blazhko' variable, V79 in M3, has  exhibited mixed mode
pulsations, we consider the first possibility to be more feasible.

It appears that regular monitoring of stars like V79 in M3 can provide
important information about stellar evolution. If there are any available
unpublished 
observations of this star between 1962 and 1996, it would be very
informative to analyze them. It would be particularly interesting to know
if there has been any change in luminosity or color because this may put
constraints on the boundaries of the different pulsational modes in the
HR diagram.  We should also point out that
there are other interesting  stars in M3 that merit
further investigation.
For example, NC noted that the RRab star V28 
was a promising RRd candidate. Unfortunately, it was not in the field 
of view for our present study. Perhaps  V28 has also started to switch modes
or will do so in the near future. It would be interesting to find out.

%
%************************ SECTION 3
%
\section{Evidence for evolution in other RRd stars}

We have presented evidence that indicates that two RRd stars in M3 are evolving
blueward. Are the RRd stars in Oosterhoff type II
systems evolving redward? This question was addressed by 
Purdue \it et al. \rm (1995) 
in a study of the long term behavior of RRd stars in the Oosterhoff type II
cluster M15. They noted that the fundamental mode
oscillations in V30 increased in strength relative to the first overtone
between 1941 and 1991. In fact, their analysis suggests that the change 
happened rather abruptly in the 1950s.
A similar situation occurs for the field RRd star, AQ Leo, which has periods
and a period ratio
similar to those of RRd stars in Oosterhoff type II systems.
Jerzykiewicz \it et al. \rm (1982) compared the amplitude differences
between a series of observations made in 1960-1961 and another
in 1973-1974 and found
that the first overtone amplitude decreased by an amount $0.012\pm 0.011$
while the fundamental mode amplitude increased by $0.017\pm 0.011$. 
They also found that an abrupt increase in the first overtone period 
occurred in the early 1970s. They stated that their findings did not 
give direct evidence for mode switching, but that it is a possibility. The
increase in strength of the fundamental mode oscillations and the period
increase are events that are expected to occur
if the star is evolving redward.

Our investigation has shown that high quality photometry of RR Lyrae
stars, particularly RRd stars and
stars that exhibit the Blazhko effect 
can provide useful information
about the evolution of horizontal branch stars. 
We expect that future studies will 
confirm this.

%
%********************** ACKNOWLEDGEMENT
%
\acknowledgements

This work has been
supported by the Natural Sciences and Engineering Research Council 
of Canada through operating grants to CMC and SMR.
RWH thanks the U.~K. Particle Physics and Astronomy Research Council for the 
award of a research grant. JK was supported by
Polish KBN grant 2P03D011.12 and by the NSF grant
AST 9528096 to Bohdan Paczy\'nski.

\newpage

%\centerline{\bf{FIGURE CAPTIONS}}

\begin{figure}
\plotfiddle{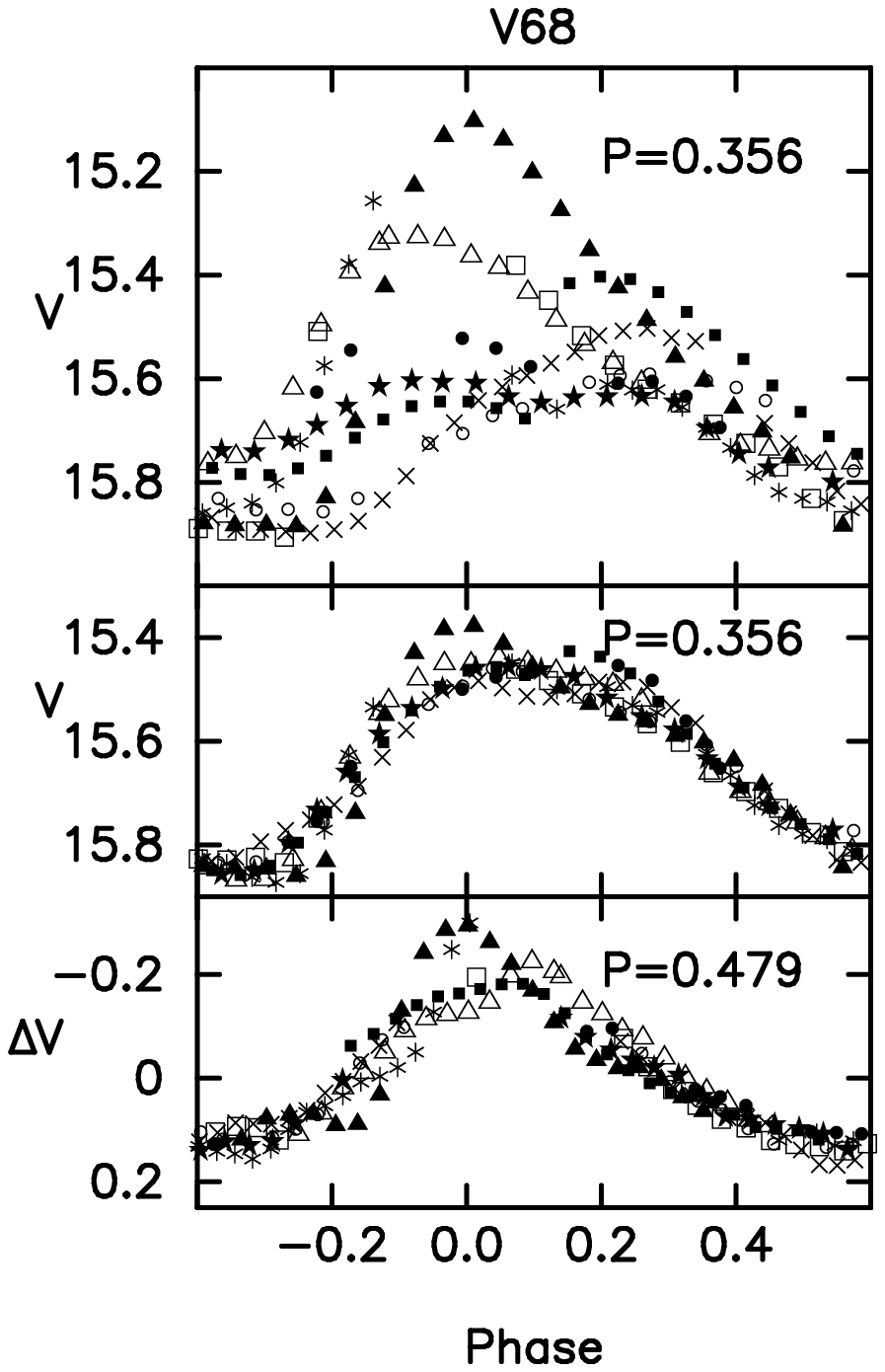}{4.0truein}{0}{80}{80}{-154}{-100}
\caption {Light curves for V68 based on the 1996 observations. In the upper 
panel, the raw magnitudes have been plotted with the dominant period, the first
overtone. 
In the middle panel, the corrected (prewhitened) magnitudes
have been plotted with the overtone period 
and in the lower panel, the (prewhitened) residuals have been plotted
with the fundamental period. 
The different symbols denote observations made on different
nights. The large night-to-night scatter of the points in the upper panel
is typical for a double-mode pulsator. The spread of the
points on curves in the lower panels is much less pronounced because
after prewhitening, the  scatter is reduced.}
\end{figure}

\newpage
\begin{figure}
\plotfiddle{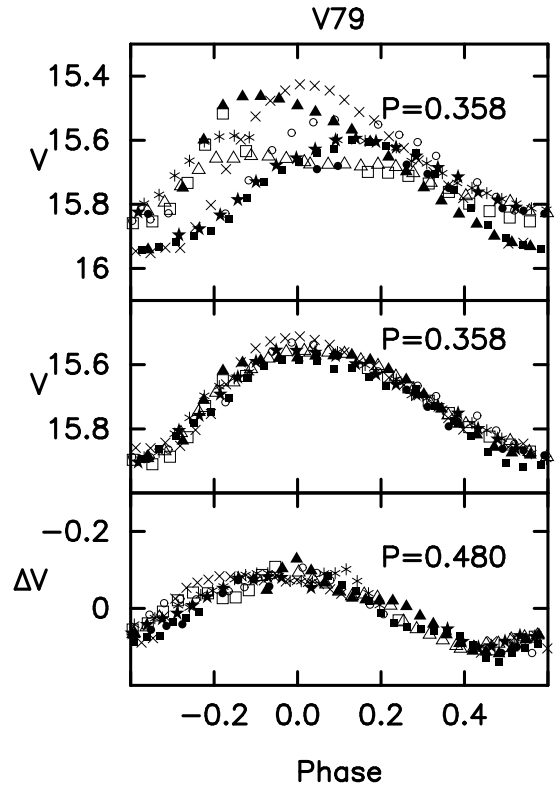}{4.0truein}{0}{80}{80}{-154}{-100}
\caption{Light curves for V79 based on the 1996 observations. The arrangement 
of the curves is the same as in Fig. 1.}
\end{figure}

\newpage
\begin{figure}
\plotfiddle{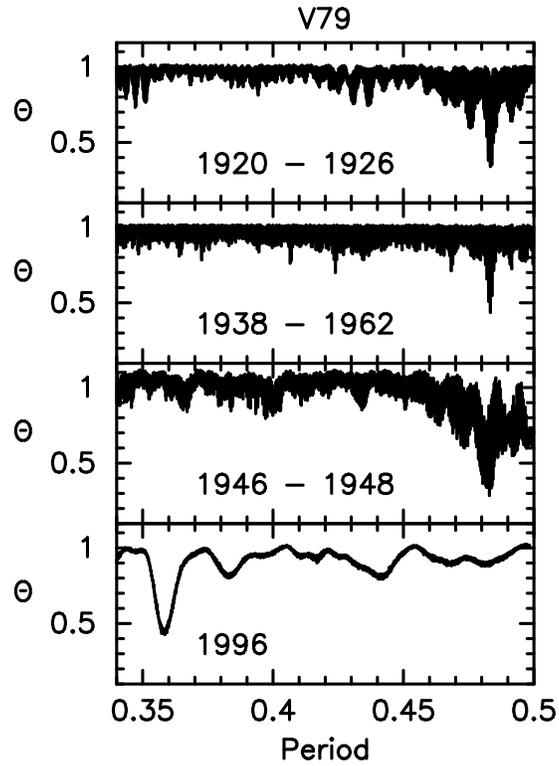}{4.0truein}{0}{80}{80}{-154}{-30}
\caption{The $\Theta$ transforms (plots of Stellingwerf's $\Theta$ statistic 
versus 
period) for V79, for four different epochs. The period for which $\Theta$ is
a minimum is considered to be the best period. The diagram illustrates
that for observations prior to 1962, the best period was approximately 
0\day 48, but in 1996, 
it was 0\day 36. This indicates a mode switch from the
fundamental to the first overtone.}
\end{figure}

%
%***********************TABLES
%
\newpage
\begin{table}
\caption{Derived properties for the RRd stars in M3 \label{Table 1}}
\begin{flushleft}
\begin{tabular}{ccccccccc}
\tableline
\tableline
Star &  $x''$ & $y''$ & $P_1$ & $P_0$ & $A_1$  & $A_0$  & $A_1/A_0$ & $A_1/A_0$ 
\cr
     &       &      &       &       & (1996) & (1996) &  (1996)   &  (1920--6) 
\cr
\tableline

V68 & +21.9 & +174.8 & 0.356 & 0.479 & 0.389  & 0.348 & $1.12\pm 0.10$ & 
$0.72\pm 0.20$ \cr
V79 & +43.4 & +349.4 & 0.358 & 0.480 & 0.337  & 0.195 & $1.73\pm 0.13$ & 
$< 0.2$ \cr
\tableline
\tableline
\end{tabular}
\end{flushleft}
\end{table} 

\end{document}